# Enhanced local viscosity around colloidal nanoparticles probed by Equilibrium Molecular Dynamics Simulations

Reza Rabani [a], Mohammad Hassan Saidi [a,*], Laurent Joly [b,c], Samy Merabia [b], Ali Rajabpour [d,†]

[a] *Center of Excellence in Energy Conversion (CEEC), School of Mechanical Engineering, Sharif University of Technology, Tehran 11155-9567, Iran*

[b] *Univ Lyon, Univ Claude Bernard Lyon 1, CNRS, Institut Lumière Matière, F-69622, Villeurbanne, France*

[c] *Institut Universitaire de France (IUF), 1 rue Descartes, 75005 Paris, France*

[d] *Advanced Simulation and Computing Laboratory (ASCL), Mechanical Engineering Department, Imam Khomeini International University, Qazvin, Iran*

**ABSTRACT:**

Nanofluids; dispersions of nanometer-sized particles in a liquid medium; have been proposed for a wide variety of thermal management applications. It is known that a solid-like nanolayer of liquid of typical thickness 0.5-1 nm surrounding the colloidal nanoparticles can act as a thermal bridge between the nanoparticle and the bulk liquid. Yet, its effect on the nanofluid viscosity has not been elucidated so far. In this article, we compute the local viscosity of the nanolayer using equilibrium molecular dynamics based on the Green-Kubo formula. We first assess the validity of the method to predict the viscosity locally. We apply this methodology to the calculation of the local viscosity in the immediate vicinity of a metallic nanoparticle for a wide range of solid-liquid interaction strength, where a nanolayer of thickness 1 nm is observed as a result of the interaction with the nanoparticle. The viscosity of the nanolayer, which is found to be higher than its corresponding bulk value, is directly dependent on the solid-liquid interaction strength. We discuss the origin of this viscosity enhancement and show that the liquid density increment alone cannot explain the values of the viscosity observed. Rather, we suggest that the solid-like structure of the distribution of the liquid atoms in the vicinity of the nanoparticle contributes to the nanolayer viscosity enhancement. Finally, we observe a failure of the Stokes-Einstein relation between viscosity and diffusion close to the wall, depending on the liquid-solid interaction strength, which we rationalize in terms of hydrodynamic slip.

---

[*] Email: saman@sharif.edu.ir (M.H.Saidi)
[†] Email: rajabpour@eng.ikiu.ac.ir (A. Rajabpour)

## I. INTRODUCTION

Nanofluids are colloidal solutions in which suspended nanometer-sized particles in the liquid medium[1–3] are responsible for enhanced thermophysical properties such as density, thermal conductivity, diffusivity, and viscosity compared to the base fluids[4–9]. The main interest behind introducing the nanofluid concept is the increment in the thermal conductivity compared to the conventional fluids[10–16]. The thermal conductivity of the nanofluids depends on various factors such as Brownian motion[17–19], thermophoresis[20–22], and clustering[23–25] of the solid particles as well as phonon-based nature of heat transfer within solid nanoparticles[26,27]. Besides, liquid molecules form a local structured layer around the nanoparticle, known as nanolayer, which acts as a thermal bridge between the nanoparticle surface and the bulk liquid[28] and affects the thermal conductivity of the nanofluids considerably. The reported values for the thickness of the nanolayer vary between the $0.3\ nm$ to $1\ nm$ depending on the applied simulation parameters[29–32]. Numerous studies have attempted to model the thermophysical properties of nanofluid considering the role of this ordered nanolayer[33–35]. As an example, the role of nanolayer in the enhanced thermal conductivity of nanofluids was investigated thoroughly in a renovated Maxwell model[36] for spherical particles suspensions and a renovated Hamilton–Crosser model[37] for non-spherical particles suspensions was developed.

The effect of nanolayer on the density of nanofluid has been also investigated experimentally and it was shown that in order to reproduce the experimental densities of the nanofluids, the nanolayer should be taken into the account[38]. Molecular dynamics (MD) simulations of silver-water nanofluids were also conducted in order to find the effect of nanolayer on the density of the nanofluid and a new ternary mixture model was proposed for the computation of nanofluids density [31]. It was shown that the existence of the nanolayer causes a contraction in the base fluid, which consequently reduces the volume of the base fluid.

Considering the viscosity of nanofluids, a few models have been proposed during the past decades [39]. For a dilute suspension of small, rigid, spherical particles in the base fluid, a good approximation of the nanofluid viscosity can be obtained by Einstein model[40]. For a moderate particle concentration, Brinkman has modified Einstein's equation and proposed a new model[41]. Batchelor[42] proposed a new model to account for the effect of Brownian motion on the viscosity of an isotropic suspension of spherical, rigid particles. There are only a few models that include the nanolayer effect in the viscosity of the nanofluid. Considering an elliptical liquid layer shell around the nanoparticles, the Ward model for calculating the viscosity was proposed[43]. The above-mentioned models for the viscosity of the nanofluid were compared with MD simulations[31]. It was shown that the calculated viscosity based on the models which do not include the nanolayer effect is not consistent with MD while the models which account for a nanolayer shell around nanoparticle follow the MD simulations viscosity well. The key role of the nanolayer viscosity highlighted in this work makes it critical to characterize it for a variety of liquid-solid interfaces. Therefore, due to the importance of the precise viscosity calculation in the nanofluid, the objective of the present research is to characterize the viscosity of the nanolayer around the nanoparticles using equilibrium molecular dynamics (EMD) simulations.

In EMD, the viscosity of the system is computed with the Green-Kubo (GK) relation. The ability of the GK relation for the calculation of bulk and confined liquid viscosity were investigated by several researchers[44–52]. The viscosity of confined water inside hydrophilic and hydrophobic nanotubes was computed using the GK method[53]. A great impact of water-wall interactions, confinement size, and density on the viscosity was observed. However, it was shown that for the small nanotube, as the density is increased, the viscosity and the diffusion coefficient are increased together and violates the Stokes-Einstein relation. Zaragoza *et al.*[54] studied how the planar walls and nanotube confinements affect water viscosity and proposed a confined Stokes-Einstein relation for obtaining the viscosity from diffusivity. It is interesting to note that, while the viscosity computed with the GK method differs from that of the bulk noticeably, the viscosity computed with confined Stokes-Einstein relation is not affected by the confinement. The authors concluded that different methods in calculating the viscosity may provide dramatically different results, which may not easily be related to the standard and experimental definition of viscosity. However, a NEMD simulation of shear-driven liquid argon flow between two graphene walls shows an increment by the factor two for the viscosity of liquid argon in the vicinity of the nanochannel wall compared to the middle of the channel[55]. Recently, Zhou *et al.*[56] have shown that in the viscosity calculation of the nanoconfined fluid, the wall friction should be decoupled from the fluid viscosity by defining the frictional (near the wall) region and the viscous (far from the wall) region. This was concluded by a comparison between the fluid viscosities calculated from the GK formula in the entire domain and the viscous region and the fluid viscosity which was obtained based on the velocity profile of the Hagen–Poiseuille flow. This analogy revealed that only the viscosity in the viscous region coincides with the one deduced from the velocity profile. The nanolayer definition coincides with the definition of the frictional region, whose viscosity was not discussed in previous studies. A closer look at the nanolayer viscosity is the main motivation of this study. Therefore, we should be able to apply the GK method to the separate liquid layers and as the first step of our research, this feasibility is validated. In addition, the diffusion coefficient of the liquid layers is also calculated, which is then connected to the difference between the viscosity of the different layers. The remainder of this article is organized as follows: In Section II, the methodology of the simulation is described in detail. In Section III, the simulation results are presented. In the first part of this section, a validation is presented to assess accuracy of the applied simulation method. Then the viscosity of the nanolayer of the copper-argon nanofluid is calculated and the results are thoroughly interpreted. Finally, we conclude in Section IV.

## II. COMPUTATIONAL DETAILS AND METHODS

In this study, the viscosity of the liquid argon is calculated by EMD method using LAMMPS[57] (Large-Scale Atomic/Molecular Massively Parallel Simulator). The Open Visualization Tool (OVITO) is used to visualize and represent atomic configurations[58]. Lennard-Jones (LJ) potential parametrization is considered for interatomic forces between different types of atoms in this study. The Lennard-Jones (LJ) potential parameters that were used in this study are shown in Table 1, in which $r_C^{LJ}$ denotes the cut-off distance of the LJ interactions. While a timestep of $1\ fs$ is used in all simulations, smaller timesteps were also tested and had no significant effect on the results. Considering the temperature of the liquid medium and the walls, a Maxwell–Boltzmann velocity distribution is used to initialize the velocity of the atoms at the beginning of the simulation. In all sets of simulations, a Nosé-Hoover thermostat was applied to the system for $2\ ns$ to ensure the desired temperature is achieved. After that, the thermostat is removed from the liquid domain while still applied on the walls. Then, the simulation proceeds for $2\ ns$. This ensured us that no thermostat is applied to the fluid atoms directly and the results are not affected by the selection of thermostat parameters. Additionally, another 2 $ns$ are performed for the averaging process in which the density, temperature, and viscosity of each computational bin are calculated.

The shear viscosity of a fluid, $\mu$, can be computed using the GK relation which is based on the shear stress $\tau_{\alpha\beta}$ autocorrelation function [59,60]:

$$\mu = \frac{V}{k_B T} \int_0^{+\infty} \langle \tau_{\alpha\beta}(t)\tau_{\alpha\beta}(0) \rangle dt, \tag{1}$$

with $\alpha \neq \beta$,

$$\tau_{\alpha\beta} = \frac{1}{V}\left( \sum_{i=1}^{N} \frac{P_\alpha^i P_\beta^i}{m^i} + \sum_{i=1}^{N} \sum_{j>i}^{N} r_{ij}^\alpha F_{ij}^\beta \right), \tag{2}$$

where $V$ and $T$ are the volume and temperature of the system, respectively, $k_B$ the Boltzmann constant, $m$ the mass, $P$ the momentum, as well as $r$ and $F$, are the distance and the force between two particles, respectively; $i$ and $j$ are the atomic indexes and $N$ is the number of particles. According to the GK relation, the calculated shear viscosity of equation (1) consists of three components, i.e., $\mu_{xy}$, $\mu_{xz}$ and $\mu_{yz}$ and the shear viscosity of the fluid equals their average.

In order to interpret the result, the diffusion coefficient was also determined using the particle's mean square displacement[54]:

$$D = \lim_{t \to \infty} \frac{\langle |r(t_0 + t) - r(t_0)|^2 \rangle}{2\, dim \times t} \tag{3}$$

where r denotes the position vector. The standard formula considering the three-dimensional displacement uses dim=3 while for two-dimensional displacement, dim=2 should be used. However, since in the current study the MSD is calculated separately in each direction and the diffusion coefficients in perpendicular and parallel directions are computed based on them, dim=1 is considered in equation (3). In order to have a reasonable estimate of the intrinsic diffusion coefficient, the liquid center of mass motion was subtracted before computing the MSD[61]. The validity of the Stokes-Einstein relation was investigated using [62–64]:

$$\mu = \frac{k_B T}{3\pi D \sigma_h} \tag{4}$$

which relates the viscosity and the diffusivity of the fluid. In this equation, $\sigma_h$ denotes the hydrodynamic fluid particle diameter. The above-mentioned equations for calculating the viscosity and diffusion coefficient are for the calculation of the entire volume of the liquid. However, as it comes to layers of liquids, these relations must be used with caution. In particular, since the calculation of the diffusion coefficient is based on the MSD according to equation (3), some restrictions must be considered. Actually, the displacement of an atom is calculated based on its reference position which is normally the original position at the onset of the MSD calculations[57]. Consequently, as an atom moves to the adjacent liquid layer, the validity of the MSD calculations become under question. In order to handle such restriction in our study, the MSD calculation time is chosen smaller than the required time that an atom needs to move to its adjacent liquid layer. Considering the diffusion coefficient of bulk liquid argon[65] and liquid layer thickness, the required time for an atom to move to its adjacent liquid layer is on the order of $0.3\ ns$ and in order to be cautious in this study, $0.2\ ns$ is considered as the maximum MSD calculation time.

**TABLE 1.** LJ parameters for non-bonding interactions

| $interaction$ | $\varepsilon\ (kcal/mol)$ | $\sigma$ (Å) | $r_C^{LJ}$ (Å) |
|---|---|---|---|
| $Ar - Ar$ [66–69] | 0.237 | 3.405 | 27 |
| $Cu - Cu$ [70] | 4.72 | 2.616 | 10 |
| $Ar - Cu$ [71,72] | 0.25, 0.5, 1.0, 1.5 | 3.010 | 18 |

### III. RESULTS

It should be noted that the GK method is derived for a bulk system [73,74] and it was verified that for a system with geometrical confinement, in the direction without confinement where the periodic boundary conditions are applied, calculated thermal conductivity based on the GK equation yields results which agree well with the Non-Equilibrium Molecular Dynamics (NEMD)[75]. Therefore, the first step in this paper is to validate the use of GK viscosity calculations for a system with a non-periodic boundary condition.

## A.  Validation of the GK methodology for the liquid layer

The simulation domain for the validation is presented in FIG. 1. The system is a cube with sides of $6\ nm$ filled with 4280 argon atoms at the temperature of $100\ K$. This corresponds to a density of $1314\ kg/m^3$. Considering the density and temperature, the viscosity is expected to be $184\ \mu Pa.s$[76]. Periodic boundary conditions are applied in x, y, and z directions. Regarding this configuration, two sets of simulations were conducted. In the first set of simulations, the viscosity of the whole simulation box is considered as a unit configuration, and the components of the viscosity for the whole simulation box are calculated using equations (1) and (2). In the second set of simulations performed simultaneously, the simulation box is divided into two equal simulation domains and the components of the viscosity in each region are calculated separately. The calculated viscosity is averaged over twenty-five different initial distribution configurations. It is observed that as the GK method is applied on the whole simulation box, the computed viscosity, $180 \pm 7\ \mu Pa.s$, is close to the expected value while as the GK method is applied on the left or the right box, the calculated viscosity becomes $130 \pm 5\ \mu Pa.s$ and $127 \pm 5\ \mu Pa.s$ respectively where the deviation from the desired value is considerable. To find out the source of such deviations, the autocorrelation of the shear stress components, averaged over all cases, is shown in FIG. 2. This figure clearly implies that the viscosity component in the $yz$ plane should be higher than in the $xy$ and $xz$ planes. The viscosity components for the left and right boxes are computed based on these autocorrelations. The calculated viscosity components for $xy$, $xz$, and $yz$ planes are $111 \pm 7\ \mu Pa.s$, $112 \pm 8\ \mu Pa.s$, and $178 \pm 8\ \mu Pa.s$ respectively for the left simulation box and $108 \pm 8\ \mu Pa.s$, $108 \pm 8\ \mu Pa.s$, and $175 \pm 8\ \mu Pa.s$ for the right ones. These values show that while the calculated viscosities in the $xy$ and $xz$ planes deviate significantly from $184\ \mu Pa.s$, the $yz$ component is in good agreement with the expected value. A closer look at the boundary conditions might reveal the source of such behavior.

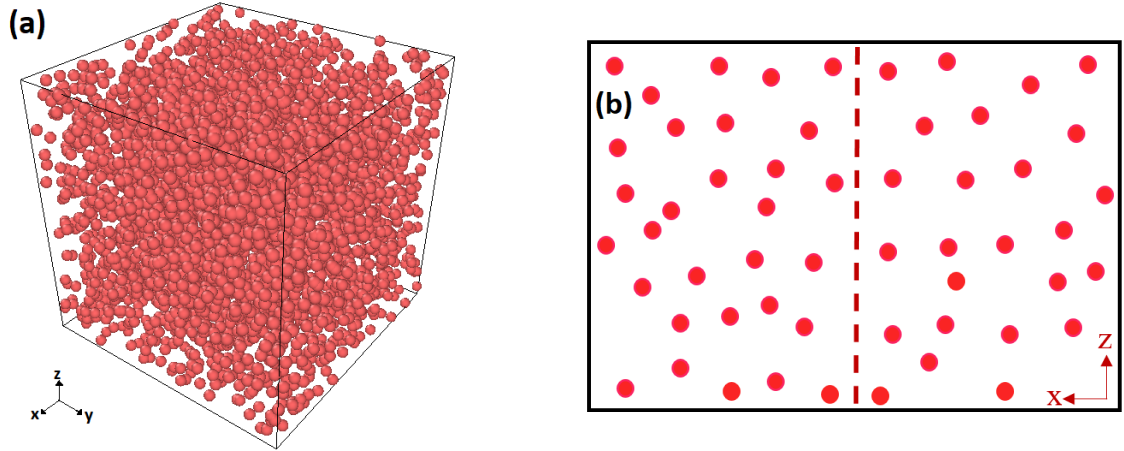

**FIG. 1.** (a) Simulation domain filled by liquid argon atoms and (b) schematic sketch of the simulation domain which is divided into two parts.

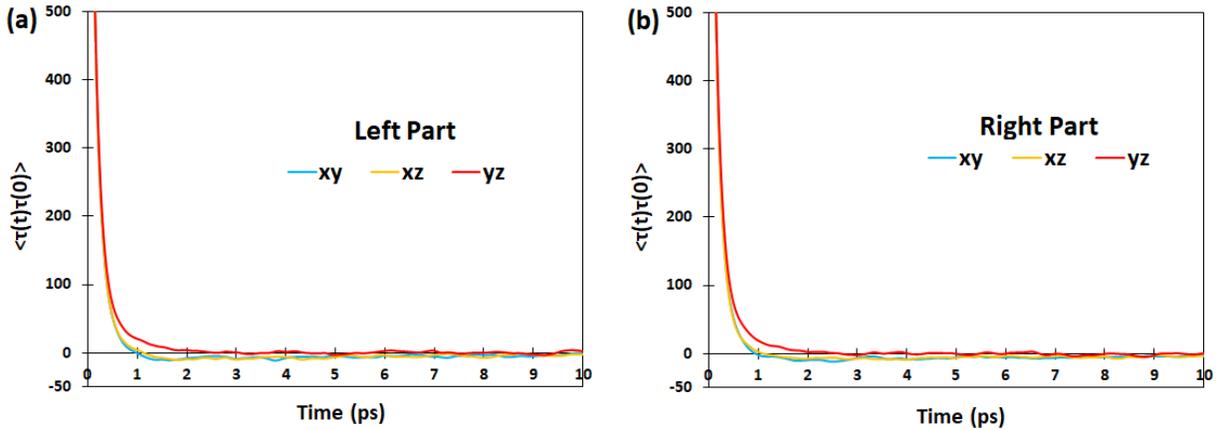

**FIG. 2.** Autocorrelation of the shear stress components for the (a) left and (b) right simulation box.

Actually, the periodic boundary condition is applied in $y$ and $z$ directions and consequently in $yz$ plane. On the other hand, the GK method underestimates the $\mu_{xy}$ and $\mu_{xz}$ due to the presence of the non-periodic boundary condition in the middle of the box along $x$ direction. It can be concluded that similar to the thermal conductivity[75], the viscosity calculation using the GK method can be only applied in a direction having periodic boundary conditions. It is also interesting to notice that the boundary condition in the middle of the box is not a wall or any physical confinement. It is just an imaginary boundary that divides the simulation box into two parts. Besides, it worth mentioning that the convergence of the autocorrelation function of all the three viscosity components does not guarantee the correctness of calculated viscosity in the GK method, and further criteria should also be checked.

### B. Copper nanoparticle in liquid argon

In the previous section, we observed that even in the presence of a non-periodic boundary condition, the viscosity component in a plane with periodic boundary conditions in all its directions still gives the local viscosity of the fluid in that region. Considering a nanoparticle and spherical nanolayer around it to calculate the viscosity, such a plane cannot be defined. For the nanoparticles with a diameter larger than 20 $nm$, a small part of the nanoparticle surface can be modeled as a flat wall approximately as can be seen in FIG. 3. Therefore, the simulation domain in this part is considered as a flat wall consisting of two layers of FCC copper atoms, which is extended 8 $nm$ in $x$ and $z$ directions, and a 5 $nm$ liquid domain. The temperature of the wall and argon is kept at 100 $K$. The liquid is divided into 5 bins of 1 nm thickness perpendicular to the wall for the averaging process in the MD method. Therefore, it should be noted that the center of the first, second, third, fourth, and fifth bins is located at 5, 15, 25, 35, and 45 nm from the wall respectively. It should be mentioned that all the presented data for the temperature, density, and viscosity is the value averaged over 20 different initial distributions of the liquid atoms. To find the effect of the interaction strength between nanoparticle surface and the liquid on the density and viscosity of the nanolayer, $\varepsilon_{Cu-Ar}$ is considered as 0.25, 0.5, 1.0 and, 1.5 $kcal/mol$. Actually, 1 $kcal/mol$ approximately equals the case with Lorentz-Berthelot mixing rules for interaction strength; $\sqrt{\varepsilon_{Cu-Cu} \times \varepsilon_{Ar-Ar}} = 1.05\ kcal/mol$. Changing the solid-liquid interaction strength varies the number of liquid atoms in the nanolayer, the first layer adjacent to the wall, which consequently affects the liquid density in the other layers. To be able to compare these solid-liquid interaction strengths together, the number of liquid atoms in the simulation domain has been changed, based on a trial-and-error method, in such a way that the density of the liquid remains approximately 1314 $kg/m^3$ for the third to the fifth layers (which can be considered as the bulk liquid). The calculated number of liquid atoms for each case is shown in Table 4.

**TABLE 4.** Calculated density of the nanolayer and its corresponding viscosity extracted from Lemmon and Jacobsen[76]

| $\varepsilon_{Cu-Ar}$ (kcal/mol) | # liquid atoms | density ($kg/m^3$) | viscosity ($\mu Pa.s$) |
|---|---|---|---|
| 0.25 | 6725 | 1367 | 217 |
| 0.5 | 6850 | 1470 | 320 |
| 1.0 | 6950 | 1540 | 426 |
| 1.5 | 7000 | 1597 | 560 |

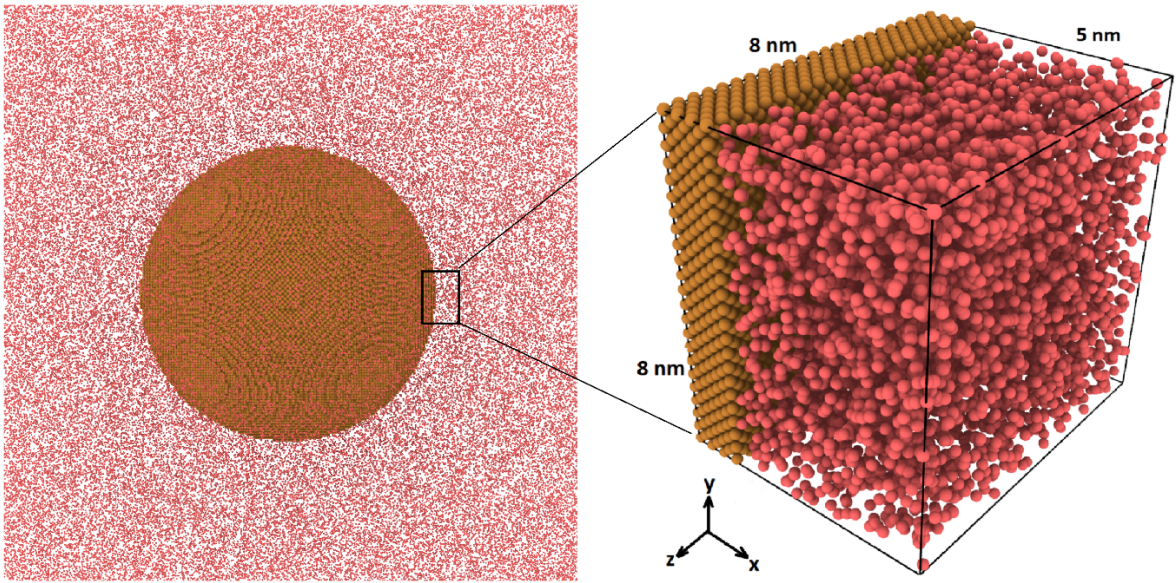

**FIG. 3.** Simulation domain considered for viscosity calculation, which is a part of 20 $nm$ diameter copper nanoparticle surface that is immersed in liquid argon.

The density distribution of the liquid in different layers is shown in FIG. 4. It can be seen that the liquid density in the first layer around the nanoparticle is enhanced notably to 1367, 1470, 1540 and 1597 $kg/m^3$ as the solid-liquid interaction strength is increased from 0.25 to 1.5 $kcal/mol$. Interestingly, the second layers also show higher densities which are 1331, 1356, 1369 and 1372 $kg/m^3$ as the solid-liquid interaction strength is increased. Meanwhile the density of the other layers is about 1314 $kg/m^3$ as expected. The increase in the density of the liquid in the first and second layers corresponds to the well-known density layering phenomenon near solid surfaces,[77] which is a direct consequence of interactions with the wall in the near-wall region[66].

As mentioned before, the spatial viscosity distribution in the liquid is calculated using the GK relation as described in the previous section. According to FIG. 3, $\mu_{xz}$ is considered as the viscosity of liquid argon in each layer. The autocorrelation function of $\tau_{xz}$ for the liquid layers is shown in the supplementary information, and the viscosity of each layer is calculated based on them. It is observed that the viscosity of the first liquid layer becomes 275, 442, 1031 and 2284 $\mu Pa.s$ as the solid-liquid interaction strength is increased from 0.25 to 1.5 $kcal/mol$ while for the second layer, the corresponding viscosities are 182, 184, 190 and 190 $\mu Pa.s$. The viscosities of the other layers are around 184 $\mu Pa.s$[76] which is the expected viscosity for 1314 $kg/m^3$. By normalizing the viscosity of each layer with respect to 184 $\mu Pa.s$, as done in Fig. 5, it is interesting to notice that the viscosity of the first layer is approximately 1.5, 2, 6, and 12 times the bulk viscosity as the interaction strength increased, which shows the great impact of the solid surface on its adjacent liquid viscosity. Interpreting the main reasons behind such anomalous viscosity increment of the first layer is the main subject of this manuscript.

One reason behind such viscosity increment is related to the denser liquid medium of the first layer compared to the other layers. A denser liquid medium increases the collision between liquid atoms which enhances the viscosity accordingly. This can be also clearly observed in the viscosity enhancement of the second layer compared to the third, fourth, and fifth layers as it has higher liquid density according to Fig. 4. It should be noted that the viscosities of the second to the fifth layer are still in the same order of magnitude which implies that a similar mechanism governs the transport phenomenon in these layers. Meanwhile, the viscosity of the first layer is much higher compared to the other layers, which suggests that a different mechanism might govern the transport phenomenon in this layer.

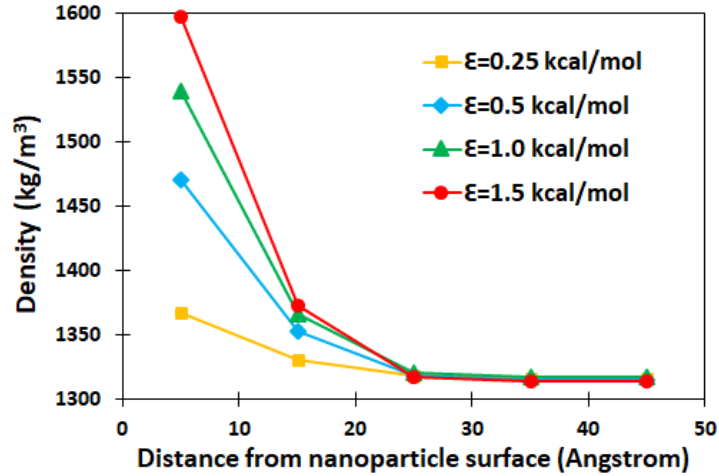

**FIG. 4.** Local liquid density in the different layers.

It might be helpful to find the viscosity of the liquid at the corresponding bulk density for the first layer. This information will show us whether the viscosity increment of the nanolayer is only related to the increase in the liquid density or if some other phenomenon is also involved. The viscosities based on the liquid density for the first layer are presented in Table 4. Lemmon and Jacobsen[76] presented these viscosities for the bulk liquid medium based on a combination of the theoretical models and the empirical equations. The ratio of this viscosity to the calculated one can be considered as the contribution of the density to the total viscosity. It is observed that the viscosity of the first liquid layer based on its density is $217, 320, 426$ and $560$ $\mu Pa.s$ as the solid-liquid interaction strength is increased from $0.25$ to $1.5$ $kcal/mol$, which shows a notable difference compared to the viscosity of the nanolayer. Therefore, it is concluded that in addition to the liquid density, there must be some other phenomenon that increase the viscosity of the nanolayer.

**TABLE 5.** Calculated viscosity ($\mu Pa.s$) for the different liquid argon layers as a function of the interaction strength between the wall and liquid atoms

| $\varepsilon_{Cu-Ar}$ ($kcal/mol$) | First | Second | Third | Fourth | Fifth |
|---|---|---|---|---|---|
| 0.25 | 275 | 182 | 173 | 172 | 174 |
| 0.5 | 442 | 184 | 175 | 173 | 174 |
| 1.0 | 1031 | 190 | 178 | 171 | 173 |
| 1.5 | 2284 | 190 | 170 | 174 | 172 |

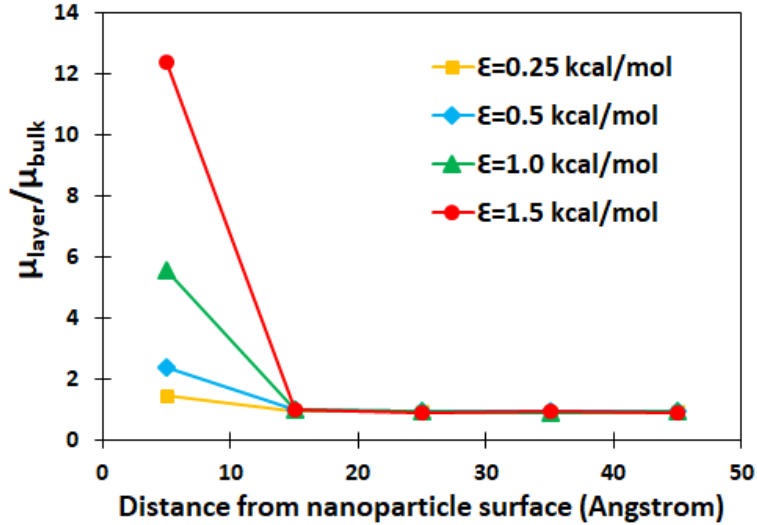

**FIG. 5.** Normalized viscosity distribution in the liquid argon normalized with $\mu_{bulk} = 184 \; \mu Pa.s$.

In the liquid nanolayer adjacent to a nanoparticle, the liquid atoms are under the strong attraction of the nanoparticle atoms on the surface. On the other hand, the reported values of Lemmon and Jacobsen[76] are for the bulk condition where the liquid atoms can move and interact with each other freely. Therefore, a closer look at the liquid structure and transport properties of the atoms in the nanolayer might reveal the main reason behind such discrepancy between the calculated viscosity of the nanolayer and the expected viscosity based on its density.

Figure 6 displays the distribution of the liquid atoms near the nanoparticle surface. As it can be observed, the liquid atoms in the nanolayer form two ordered liquid atoms layers for the interaction strength of 0.25 kcal/mol while three layers are formed for higher interaction strength. We suggest that the anomalous viscosity enhancement of the nanolayer compared to the other layers can be assigned to such liquid atoms ordering, and we will refer to it as the contribution of the liquid structure to viscosity. This can be computed by subtracting the contribution of the liquid density from total viscosity. Figure 7 compares the contribution of the liquid density and the layered structure in the total viscosity of the nanolayer which shows a greater role played by the liquid structure as the liquid-solid interaction strength increased.

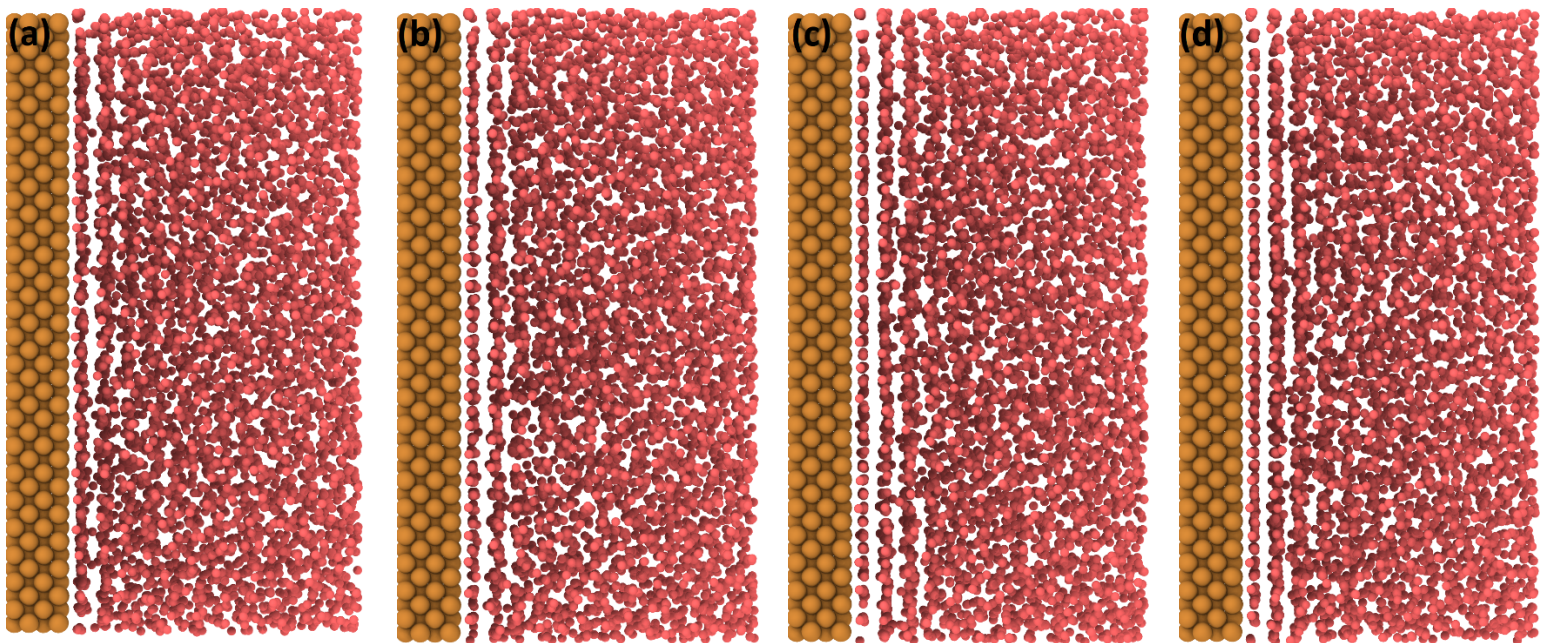

**FIG. 6.** Distribution of the liquid atoms near the nanoparticle surface in the steady state for liquid-solid interaction strength of 0.25 (a), 0.5 (b), 1.0 (c), and 1.5 kcal/mol(d).

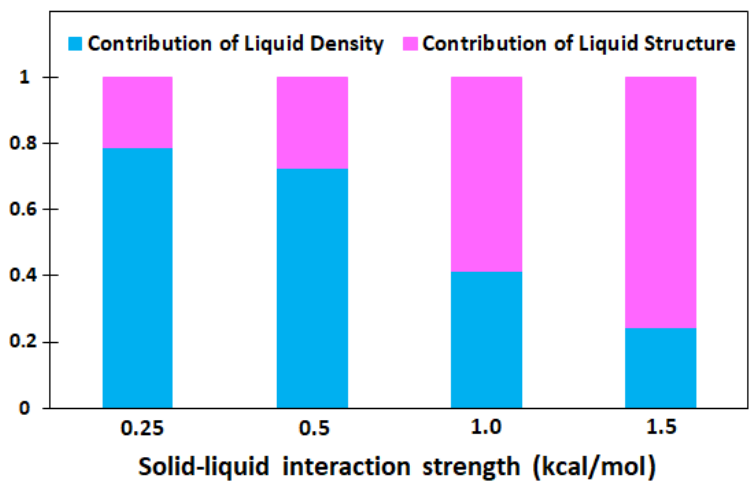

**FIG. 7.** Contribution of liquid density and liquid structure in the total viscosity for different interaction strength of solid/liquid.

To go further, the mean square displacement (MSD) and the diffusion coefficient can be considered as the main criteria of the transport characteristics of the liquid atoms. Therefore, a more detailed evaluation of these parameters might help us to have a better understanding of the underlying physics in the nanolayer. The diffusion coefficients are calculated based on the MSD of the liquid atoms using equation (3). According to Fig. 3, the diffusion coefficient perpendicular to the nanoparticle surface is $D_x$ and the average of $D_y$ and $D_z$ is considered as the diffusion coefficient parallel to the surface. Figure 8 shows the variation of the diffusion coefficient parallel and perpendicular to the nanoparticle surface as a function of solid-liquid interaction strength for liquid layers. It is observed in Fig. 8a that as the liquid-solid interaction energy increases from 0.25 to 1.5 $kcal/mol$, the diffusion coefficient parallel to the nanoparticle surface in the third to the fifth layers, is about $3.35 \times 10^{-9} m^2 s^{-1}$ while a reduction from $3.02 \times 10^{-9}$ to $2.76 \times 10^{-9} m^2 s^{-1}$ and $1.86 \times 10^{-9}$ to $0.65 \times 10^{-9} m^2 s^{-1}$ is observed for the second and first layers respectively. Figure 8b clearly shows that for the same increment of interaction energy, the diffusion coefficient perpendicular to the nanoparticle surface is approximately $3.30 \times 10^{-9} m^2 s^{-1}$ for the fourth and the fifth layers, while it decreases from $2.92 \times 10^{-9}$ to $2.72 \times 10^{-9} m^2 s^{-1}$ for the third layer. For the second and first layers, the diffusion coefficient decreases more, from $2.16 \times 10^{-9}$ to $1.83 \times 10^{-9} m^2 s^{-1}$ and $1.09 \times 10^{-9}$ to $0.5 \times 10^{-9} m^2 s^{-1}$ respectively. The diffusion coefficient perpendicular direction to the nanoparticle surface is inversely related to the residence time of the liquid atoms in the layers. Therefore, it can be concluded that the liquid atoms remain more time in the first liquid layer compared to the other layers for all interaction strength according to Fig. 8b. More details about the MSD and diffusion coefficient of the different layers are shown in the supplementary information.

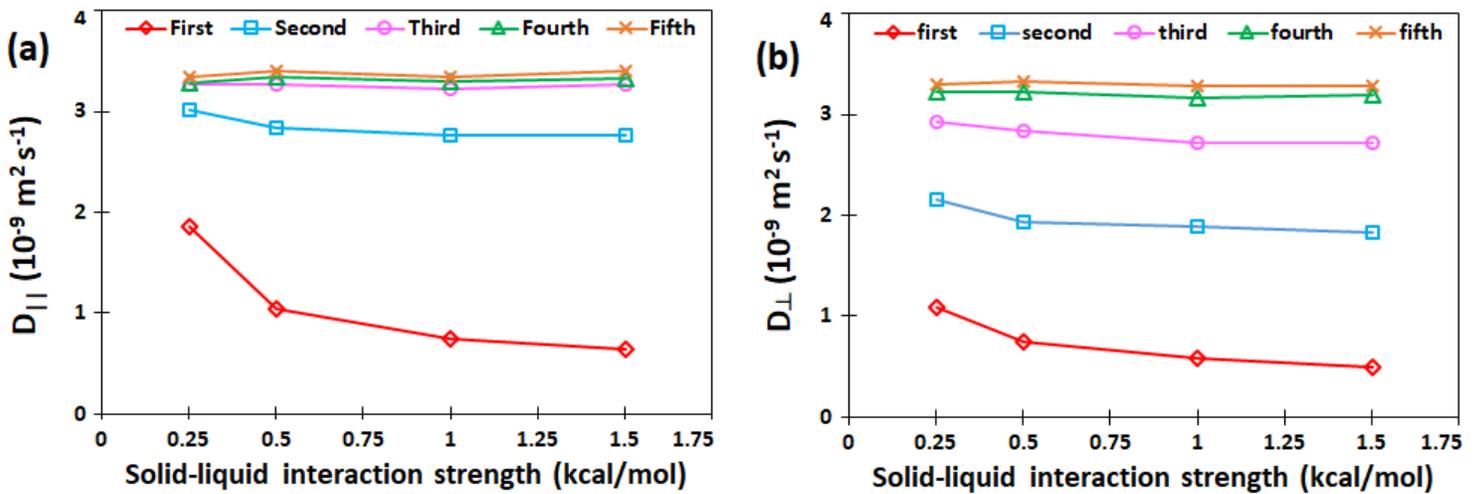

**FIG. 8.** Variation of the diffusion coefficient parallel (a) and perpendicular (b) to the nanoparticle surface as a function of solid-liquid interaction strength for liquid layers

Using equation (4), the validity of the Stokes-Einstein relation for each layer is also investigated in Fig. 9. Considering the nanolayer, it is interesting to notice that the Stokes-Einstein relation is valid for the low and intermediate solid-liquid interaction strength as $D_\parallel \times \mu$ is approximately constant for all liquid layers in Fig. 9a. However, it should also be noted that a small reduction in the first layer is observed. The Stokes-Einstein relation implies a hydrodynamic radius which should depend on the local density, but if we take into account this correction as $D_\parallel \times \mu \times \rho^{-1/3}$, we still have a small deviation for the first layer according to Fig. 9b. On the other hand, Figs. 9a and 9b clearly show that for stronger interaction, the violation of the Stokes-Einstein relation becomes noticeable. This violation is mainly related to the different structural and transport properties of the first layer compared to the others as the solid-liquid interaction strength increased.

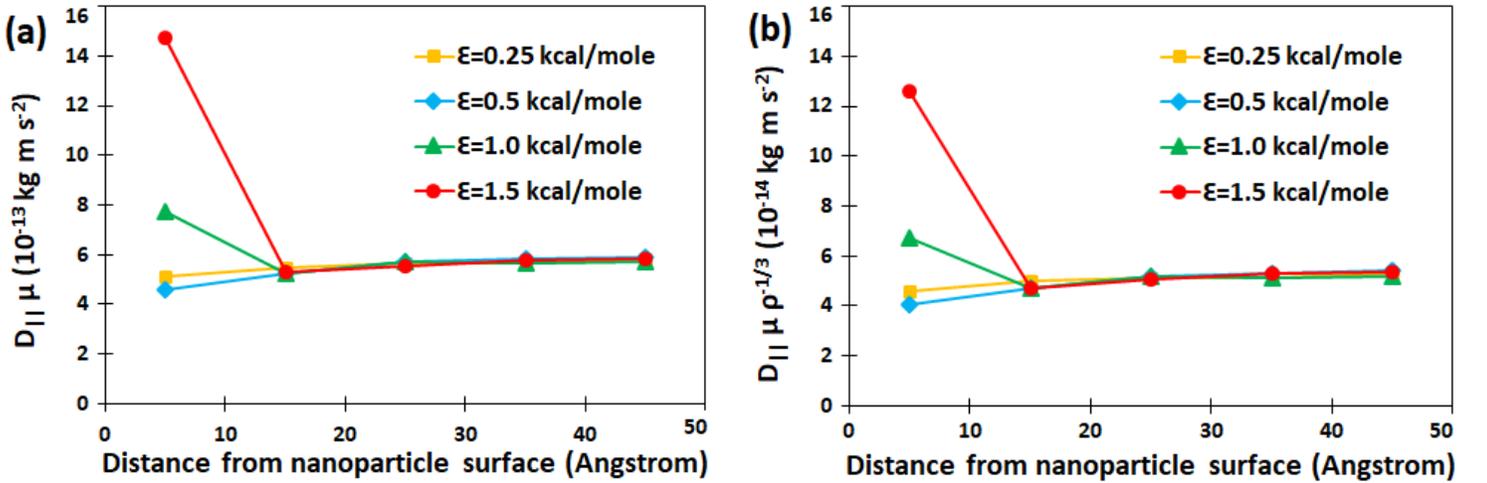

**FIG. 9.** The evolution of the Stokes-Einstein $D_\parallel \times \mu$ (a) and modified Stokes-Einstein with hydrodynamic radius $D_\parallel \times \mu \times \rho^{-1/3}$ (b) relation as a function of distance from the nanoparticle surface.

The variation of the viscosity and the diffusivity of the nanolayer against the interaction strength are shown in Fig. 10 in which the solid lines show the exponential behavior. It is observed that the viscosity and the diffusion coefficient of the nanolayer parallel to the nanoparticle surface vary as $\left[73 e^{\left(\frac{\varepsilon}{k_B T}\right)^{0.61}}\right] \times 10^{-6}\ Pa.s$ and $\left[0.8 e^{\left(\frac{\varepsilon}{k_B T}\right)^{-1.33}}\right] \times 10^{-9}\ m^2 s^{-1}$ respectively as the solid-liquid interaction strength increases.

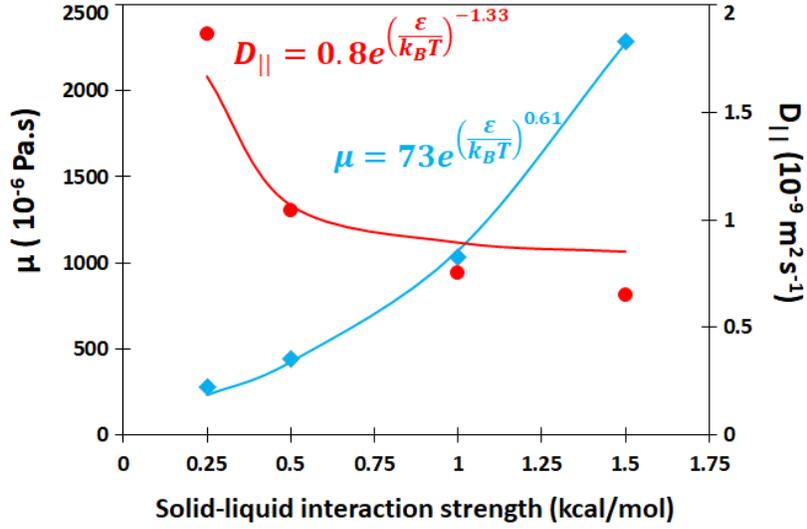

**FIG. 10.** Viscosity and diffusion of the nanolayer as a function of liquid-solid interaction strength.

The effect of liquid-solid slip on the diffusion coefficient close to the wall of the nanoconfined liquid medium was investigated thoroughly[78,79]. Saugey et al.[80] studied the effect of slip length at the solid-liquid interface on the diffusion of the confined liquid numerically, using a hydrodynamic approach based on the Stokes equation. It was shown that for the no-slip boundary condition, the diffusion coefficient in the vicinity of the wall decreased up to 50% compared with the bulk diffusion, while as the slip on the boundary increased, the diffusion coefficient was enhanced by the same amount. Interestingly, the authors have also derived an analytical expression for low and high confinement limits, which were in good agreement with numerical data. Using this analytical approach, we have proposed a similar analytical formula for our simulation domain as:

$$D_{\|}(x) = D_{bulk} \left\{ 1 - \left(\frac{9}{16}\right) \frac{\sigma}{x + b - \sigma} \right\} \qquad (5)$$

where $\sigma$ is the diameter of the liquid atoms which is 3.405 Angstrom in our case, $b$ is the slip length, and $D_{bulk}$ is the diffusion coefficient in the bulk region away from the solid surface. It should also mention that according to the simulation domain in Fig. 3, we have also considered the full formula accounting for periodic boundary conditions, and this does not lead to a better agreement.

The diffusion coefficients derived based on the MD simulation are shown in Fig. 11a along with the fitted analytical formula for different interaction energy. The bulk diffusion coefficient $D_{bulk}$ and the slip length $b$ in equation (5) are considered as the fitting parameters and their corresponding values are obtained for each interaction energy and shown in Fig. 11b. It can be observed that $D_{bulk}$ is around $3.5 \times 10^{-9}\ m^2 s^{-1}$, independently of the interaction energy. Meanwhile, Fig. 11b shows that the slip length declines from 2.51 to 0.75 Å as the interaction energy increases, as expected[81].

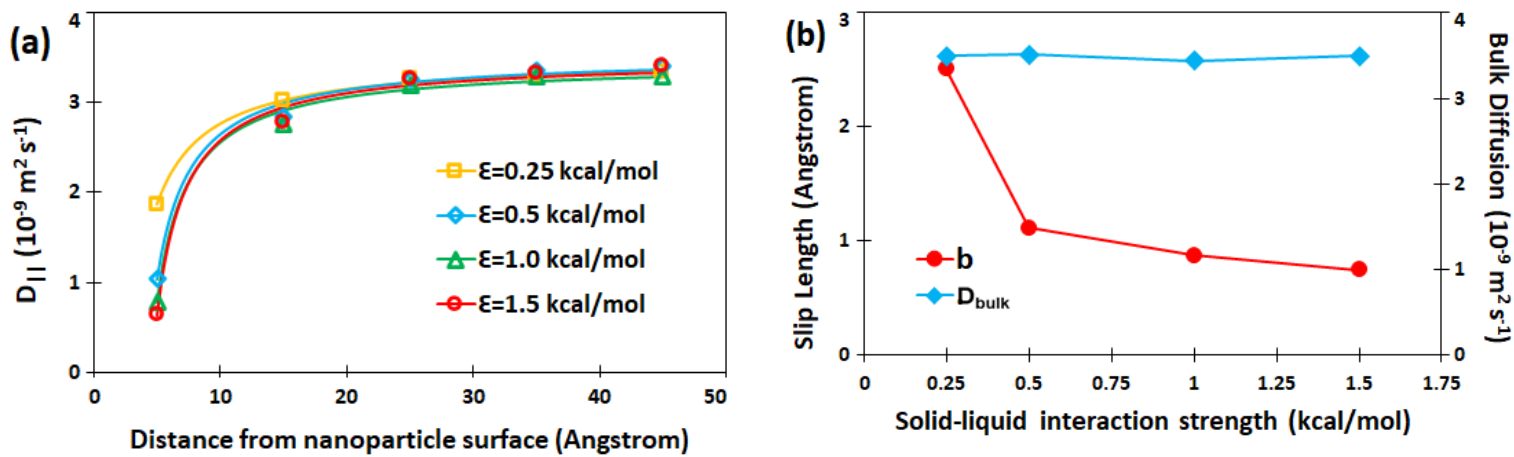

**FIG. 11.** Variation of the diffusion coefficient in different layers and its corresponding fitted analytical formula which is shown by the solid lines (a) and the variation of calculated slip length and bulk diffusion (b) as a function of solid-liquid interaction strength based on the proposed analytical formula

## IV. SUMMARY AND CONCLUSION

The EMD simulation of liquid argon around a copper nanoparticle reveals that the density increments can be observed at 2 nm from the nanoparticle surface, while it is much more intense in the first 1 nm known as the nanolayer. In this study, we characterize the corresponding viscosity increase using Green-Kubo approach. It is known that the GK method is traditionally derived for the bulk medium in which all the boundary conditions are periodic. Considering the liquid medium adjacent to the nanoparticle surface, we have shown that the GK method can be applied for such configuration with non-periodic boundary conditions provided that only the viscosity component in the normal plane to the non-periodic direction is considered in the calculation. Then, the viscosity of each layer is calculated and it was shown that for the liquid-solid interaction strength of 0.25, 0.5, 1.0 and, 1.5 kcal/mol, the viscosity is multiplied by approximately 1.5, 2, 6, and 12. We have shown that a fraction of this anomalous viscosity enhancement originates from the increment in the liquid density of the nanolayer. The further increment is suggested to be related to the ordered structure of the distribution of the liquid atoms in the nanolayer. We have shown that the contribution of the liquid structure to the total viscosity of the nanolayer is increased as the solid-liquid interaction strength is enhanced. Generally, it was inferred that for the low and intermediate interaction energy systems, the enhanced viscosity of the nanolayer can be assigned to the small increase of the local density. For these systems, the Stokes-Einstein relation -linking viscosity and diffusion- holds approximately in the nanolayer and the effect of the wall friction is small. On the other hand, for the strong interaction energy systems, the increase of the density cannot explain the enhancement of the local viscosity. This enhancement should be rather controlled by the friction with the wall and other neighboring liquid atoms. In this regime, one expects deviations to the Stokes-Einstein relation as it was observed. Finally, we rationalized the deviations from Stokes-Einstein relation by accounting for the effect of liquid-solid slip-on diffusion close to the wall, with a slip length that decreases with increasing liquid-wall interaction.

## CONFLICT OF INTEREST

There are no conflicts to declare.

## ACKNOWLEDGMENTS

R.Rabani and M.H.Saidi wish to thank the financial support by the Iran National Science Foundation (INSF). L. Joly is supported by the Institut Universitaire de France.## REFERENCES

[1] S.U.S. Choi, J. Heat Transfer **131**, 1 (2009).
[2] P. Katiyar and J.K. Singh, J. Chem. Phys. **150**, 044708 (2019).
[3] K.R.V. Subramanian, T.N. Rao, A. Balakrishnan, M.E. Zayed, S.W. Sharshir, J. Shaibo, F.A. Hammad, M.K.A. Ali, S. Sargana, K.K. Salman, E.M.A. Edreis, J. Zhao, C. Du, and A.H. Elsheikh, in *Nanofluids Their Eng. Appl.* (CRC Press, 2019), pp. 405–429.
[4] R. Taylor, S. Coulombe, T. Otanicar, P. Phelan, A. Gunawan, W. Lv, G. Rosengarten, R. Prasher, and H. Tyagi, J. Appl. Phys. **113**, 011301 (2013).19

**Supplementary**

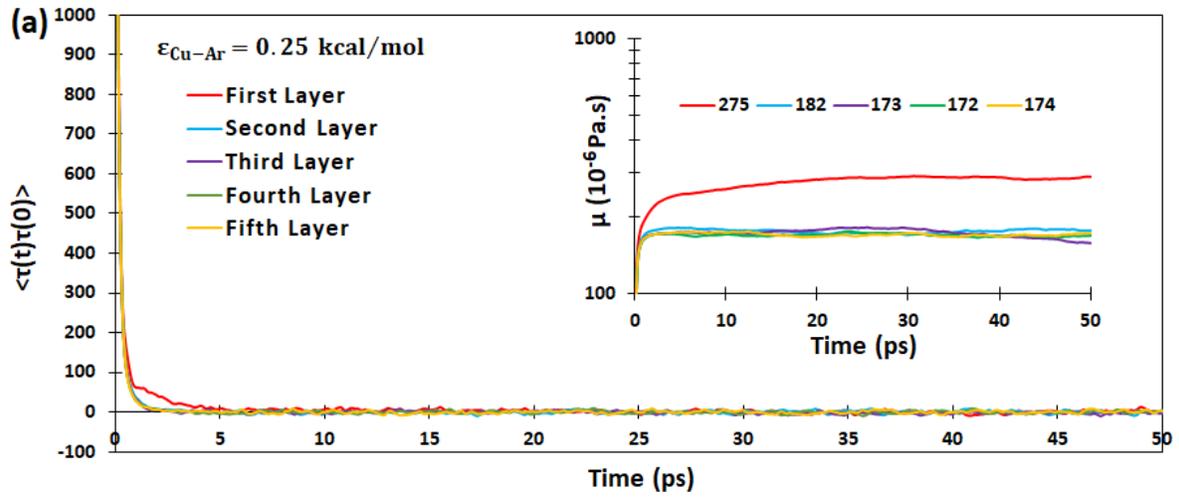

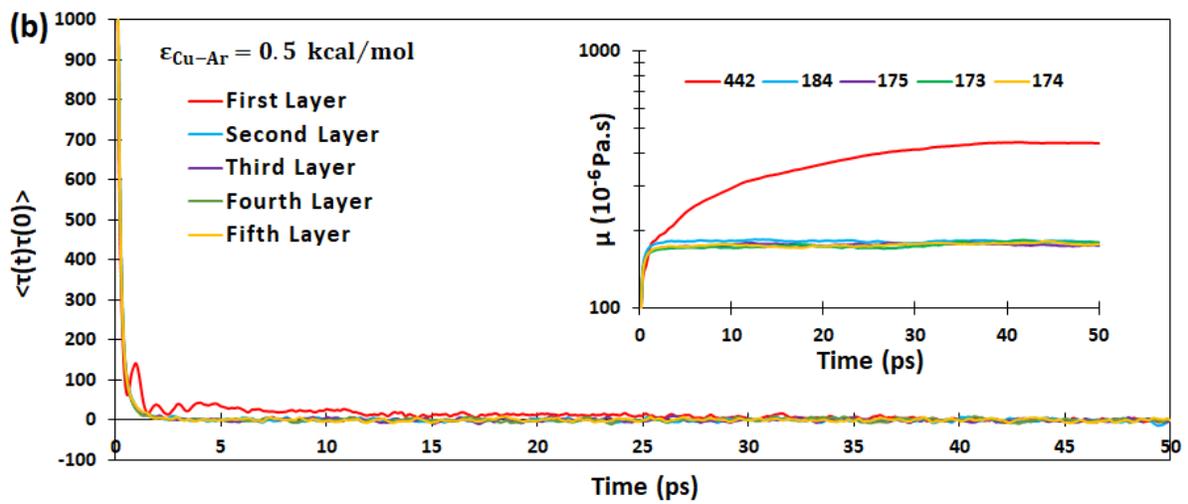

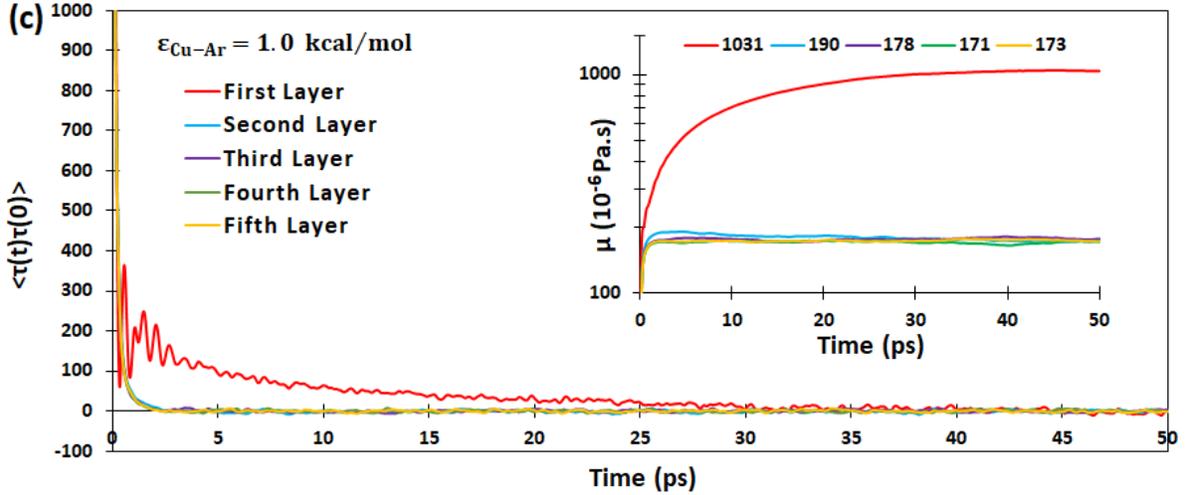

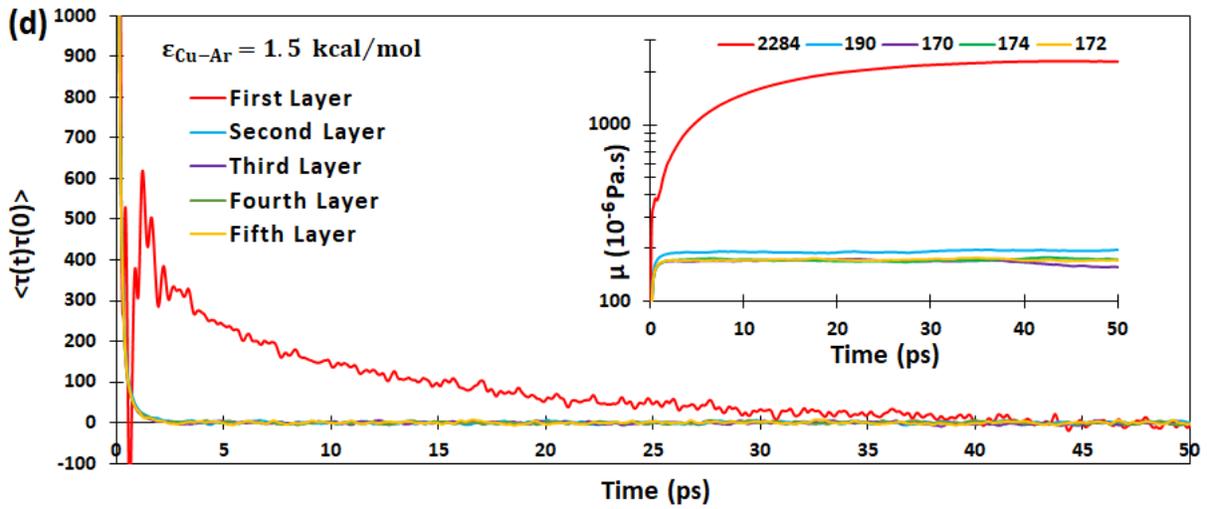

**FIG. S1.** Autocorrelation of the shear stress components in different layers of the liquid for $\varepsilon_{Cu-Ar}$ equal (a) 0.25 (b) 0.5 , (c) 1.0 , and (d) 1.5 $kcal/mol$.

Figure S1 shows the autocorrelation of the shear stress components in different layers of the liquid. A good convergence needs 50 $ps$ to be achieved. It can be seen that the autocorrelation function for the first liquid layers shows different behavior in comparison with the other layers.

Figure S2 shows the evolution of the MSD of the liquid atoms parallel to the nanoparticle surface in all five layers along with corresponding calculated diffusion coefficients parallel to the nanoparticle surface using equation (3) of the main text, for different interaction strengths. It can be observed that the MSD and the diffusion coefficients of the liquid atoms in the first layer are highly reduced compared to the other layers. Figure S2 also implies that as the interaction strength between the nanoparticle and liquid atoms increases, the reduction in the MSD intensifies.

Figure S3 displays the MSD of the liquid atoms in the x-direction perpendicular to the nanoparticle interface in all five layers along with corresponding calculated diffusion coefficients perpendicular to the nanoparticle surface using equation (3), for different interaction strengths. It shows that the MSD of the first layer of the liquid atoms in the perpendicular direction of the nanoparticle surface is affected the most by the nanoparticle force field among the other layers. Besides, as the interaction strength increased, the MSD reduction is also intensified.

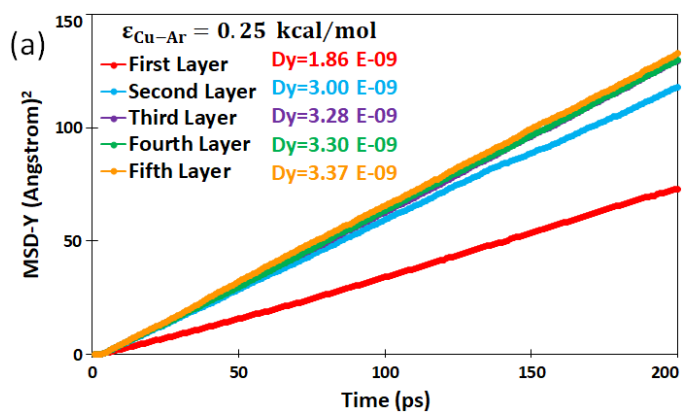
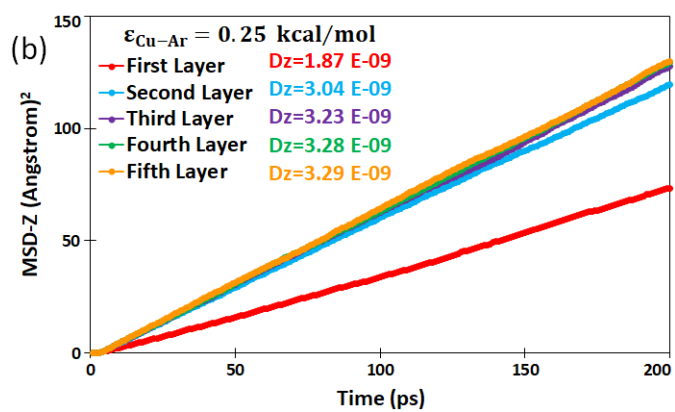
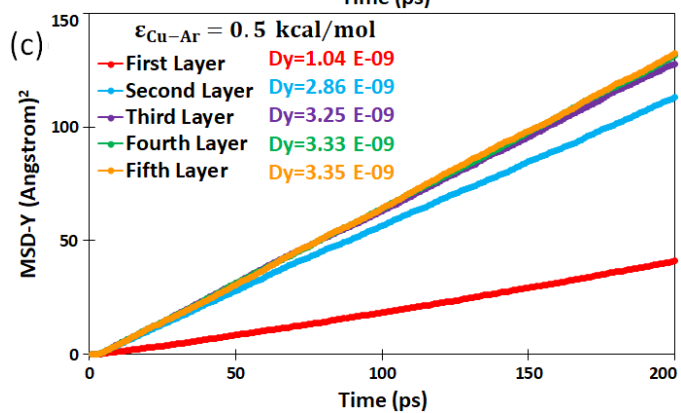
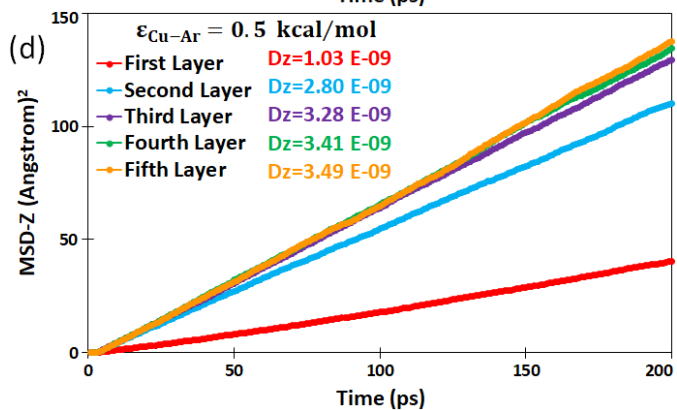

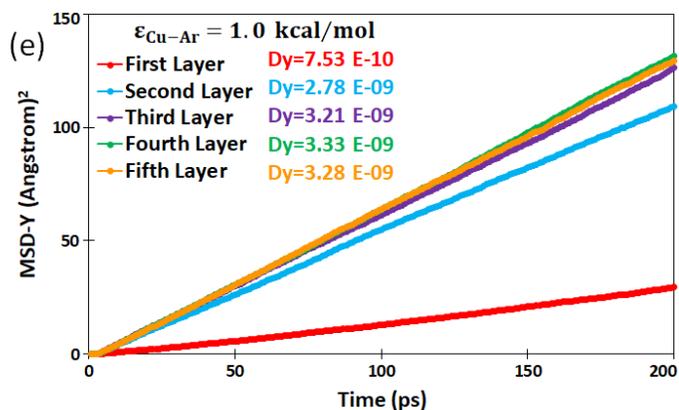
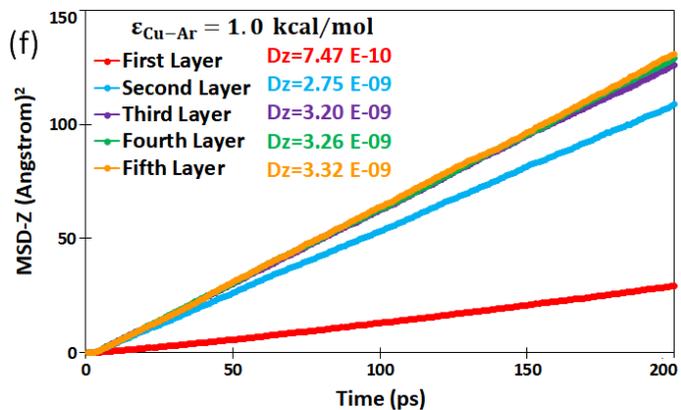
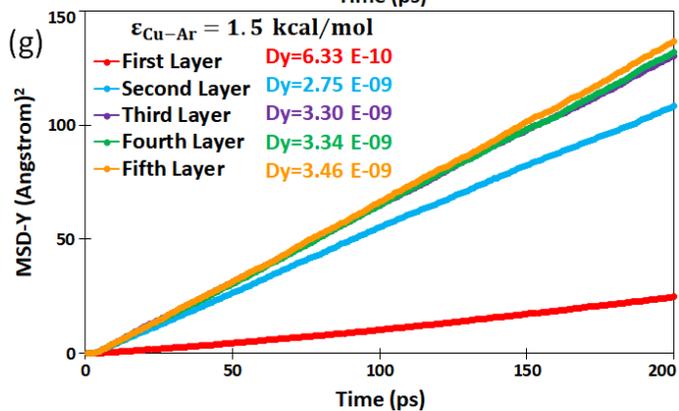
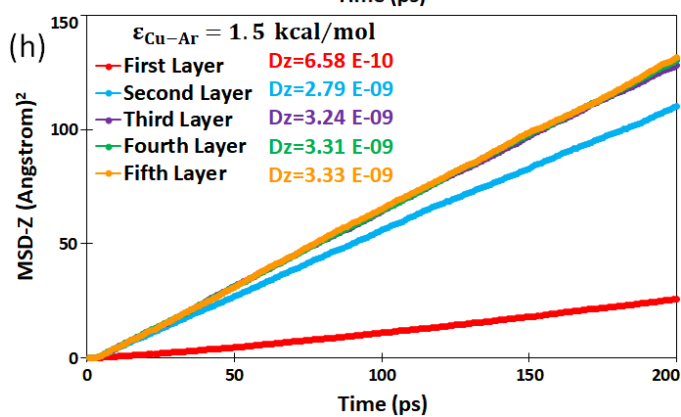

**FIG. S2.** Evolution of the MSD and the diffusion coefficient (in $m^2 s^{-1}$) of the liquid atoms in the parallel directions (y and z) to the nanoparticle surface, for different solid-liquid interaction strength.

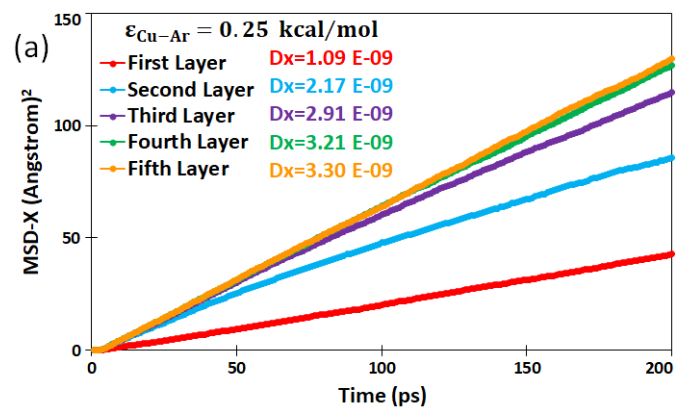
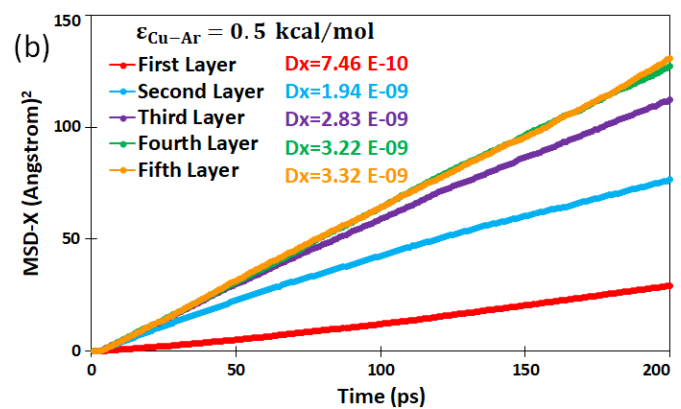

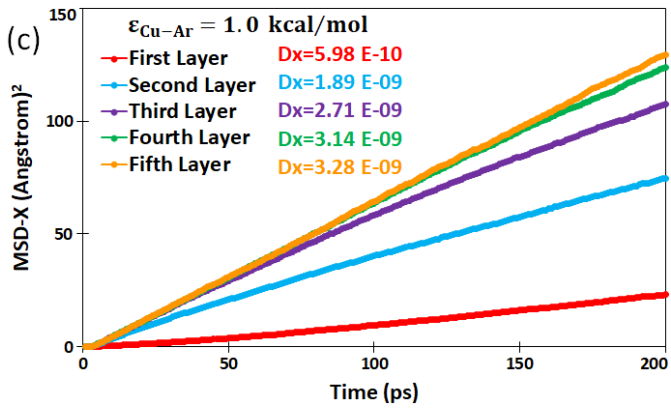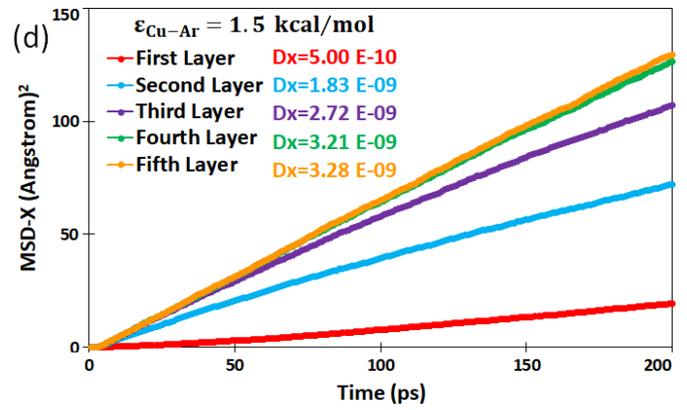

**FIG. S3.** Evolution of the MSD and the diffusion coefficient (in $m^2 s^{-1}$) of the liquid atoms in the perpendicular direction (x) to the nanoparticle surface, for different solid-liquid interaction strength.